# Theoretical study of conventional semiconductors as transducers to increase power and efficiency in betavoltaic batteries

D. Ghasemabadi*, H. Zaki Dizaji, M. Abdollahzadeh
Physics Department, Faculty of Science, Imam Hossein University, Tehran, Iran.

## Abstract

Semiconductor materials play an important role as transducers of electrical energy in betavoltaic batteries. Optimal selection of effective factors will increase the efficiency of these batteries. In this study, based on common semiconductors and relying on increasing the maximum efficiency of betavoltaic batteries and the possibility of using $^{3}H$, $^{63}Ni$, and $^{147}Pm$ beta sources, the indicators and criteria for optimal selection of semiconductor materials are determined. Evaluation criteria include the backscattering coefficient of beta particles from semiconductors, efficiency of electron-hole pairs generation, electronic specifications and properties, radiation damage threshold, radiation yield, stopping power and penetration of beta particles in semiconductors, physical characteristics, temperature tolerance, accessibility, and fabrication are considered. Conventional semiconductors have been quantitatively evaluated based on these criteria and compared with silicon semiconductors. 10 semiconductors, $\beta - B$, diamond, 2H-SiC, 3C-SiC, 4H-SiC, AlN, MgO, B4C with effective atomic number less than 14 and bandgap energy above 1.12 eV at room temperature (300K) compared to Silicon semiconductors are evaluated. Finally, according to the evaluation indicators, Diamond, c-BN, and 4H-SiC are more suitable semiconductors in terms of efficiency have selected, respectively. The results indicate that for planar batteries, a betavoltaic semiconductor type junction for Schottky diamond with $^{147}pm$ radioisotope, and 4H-SiC semiconductors with $^{63}Ni$ or $^{3}H$ radioisotopes, and for three-dimensional structures of betavoltaic batteries, Si combination with $^{147}pm$ or $^{63}Ni$ radioisotopes is recommended.

## 1- Introduction

Betavoltaic batteries are nuclear batteries that convert the energy emitted from beta-isotope sources into electrical energy using a semiconductor as a transducer. These batteries have several unique features. Such as providing voltage and current for a long time (several years to several decades), high energy density, and small size (these batteries can be reduced in size to micrometer to micrometer sizes). This superiority of betavoltaic batteries compared to other types of batteries has made them an attractive energy source to meet the future needs of electronic components.

In recent years, photovoltaic batteries have become an ideal energy source for microelectromechanical systems [1]. A betavoltaic cell consists of two main parts: the source and the semiconductor part. The connection of the semiconductor part is pn, pin, or Schottky. Parameters related to semiconductor material such as impurity concentration, the width of the depletion region, lifetime of minority carrier, connection type, energy discharge profile of beta particles in semiconductor, as well as processes including distribution, production, separation, and recombination of electron-hole pairs, in output power and conversion efficiency. Betavoltaic cell energy is effective[2]; Therefore, the optimal selection of the semiconductor as a conversion unit in the betavoltaic cell is of particular importance. However, betavoltaic batteries have been simulated and fabricated with different semiconductors[3], but the selection and using semiconductor materials have not been determined according to specific criteria and have not been quantitatively investigated.

In this research, the optimal choice of semiconductor converter has been investigated to increase power and efficiency in betavoltaic batteries. For this purpose, first, by examining the effective factors in the energy conversion efficiency of betavoltaic batteries, the selection criteria and indicators related to semiconductors are calculated and quantitatively evaluated is done. Then, according to the common and available semiconductors, the optimal semiconductor materials are determined.

## 2-Theoretical Foundations

The overall efficiency of any system is defined as the ratio of the output power to the system's input power. The total energy conversion efficiency of the betavoltaic cell ( $\eta_{total}$ ) is defined as the ratio of the maximum output power of the cell to the total power of the beta source and is obtained according to equation (1)

$$\eta_{total} = \frac{P_{out}}{P_{in}} \tag{1}$$

The total power of the beta source is according to equation (2). In this equation, q is the charge of the electron, A is the activity of the source and $E_{av\beta}$ the average energy of the spectrum of beta particles.

$$P_{in} = qAE_{av\beta} \tag{2}$$

The output power of the betavoltaic cell is obtained from equation (3).

$$P_{out} = V_{oc} I_{sc} FF \tag{3}$$

In this regard,$V_{OC}$ is open circuit voltage, $I_{SC}$ is short circuit current and FF is the filling factor, which is experimentally obtained from equation(4)[4].

$$FF = \frac{\frac{q}{K_B T} V_{oc} - Ln(\frac{q}{K_B T} V_{oc} + 0.72)}{\frac{q}{K_B T} V_{oc} + 1} \tag{4}$$

In this equation, q is the charge of the electron, T is the temperature in Kelvin, $K_B$ is Boltzmann's constant, and the open circuit voltage of the cell is obtained from equation (5)

$$V_{oc} = \frac{nK_B T}{q} \ln(1 + \frac{J_{sc}}{J_0}) \tag{5}$$

In this regard, $J_{sc}$ is the short circuit current density and $J_0$ is the saturation current density and n is the ideal coefficient for the semiconductor[5]. The saturation current in semiconductors can be obtained from equation(6)

$$J_0(A/cm^2) = 1.5 \times 10^5 \exp(-\frac{E_g}{k_B T}) \tag{6}$$

Here, $E_g$ is the energy gap in terms of electron volts, $k_B = 1.38 \times 10^{-23} JK^{-1}$ Boltzmann's constant, and T is the temperature in terms of Kelvin[6]. The maximum output current of the betavoltaic cell can be calculated from equation (7).

$$I_{\max} = \frac{qAE_{av\beta}}{w} = \frac{qAE_{av\beta}}{2.8E_g + 0.5} \tag{7}$$

Here, w is the average energy to produce an electron-hole pair. This equation shows that the energy value of the semiconductor band gap has an inverse relationship with the output current. If there is no experimental value of w for the semiconductor, Klein's formula according to equation (8) can be used to calculate it[7].

$$w = 2.8E_g + 0.5eV \tag{8}$$

Approximately, the energy required to produce an electron-hole pair is 3 times of the semiconductor bandgap. To determine the effect of different parts on the efficiency of the betavoltaic cell, the total energy conversion efficiency $\eta_{totall}$ or the overall efficiency of the betavoltaic cell can be defined as equation (9)[8].

$$\eta_{totall} = \eta_\beta \eta_{couble} \eta_{semi} \tag{9}$$

Where, $\eta_\beta$ the efficiency of the beta-phase source is the coupling efficiency $\eta_{couble}$ of the source to the semiconductor and $\eta_{semi}$ is the semiconductor efficiency. It is obvious that in order to have the maximum efficiency of the betavoltaic cell, three efficiencies, $\eta_\beta$ , $\eta_{couble}$ , $\eta_{semi}$ should be maximized.

## 2-1- Efficiency of Beta Source ($\eta_\beta$ )

$\eta_\beta$ is the fraction of electrons emitted from the source that reaches the surface of the converter. This efficiency depends on the Beta Source.

$$\eta_\beta = \frac{N_\beta}{N_0} \quad (10)$$

$N_0$ The total number of beta particles emitted from the source and $N_\beta$ the number of beta particles emitted from the source that reaches the surface of the semiconductor converter. The self-absorption effect of the radioisotope source and the directional loss due to the isotropic radiation of beta particles are the main factors in this efficiency[9]. The selected isotopes used for betavoltaic batteries in this research are according to table (1).

Table 1: Selected isotopes used in the betavoltaic battery

| **Isotope** | **Half-life (years)** | **The average energy of the spectrum (keV)** | **Maximum spectrum energy (keV)** |
|---|---|---|---|
| $^{147}$Pm | 2.62 | 61.9 | 224.6 |
| $^{63}$Ni | 100.1 | 17.4 | 66.95 |
| $^{3}$H | 12.32 | 5.7 | 18.59 |

The spectrum, average, and maximum beta energy of $^{147}$Pm, $^{63}$Ni, and $^{3}$H isotopes are shown in figures (1) to (3) [10].

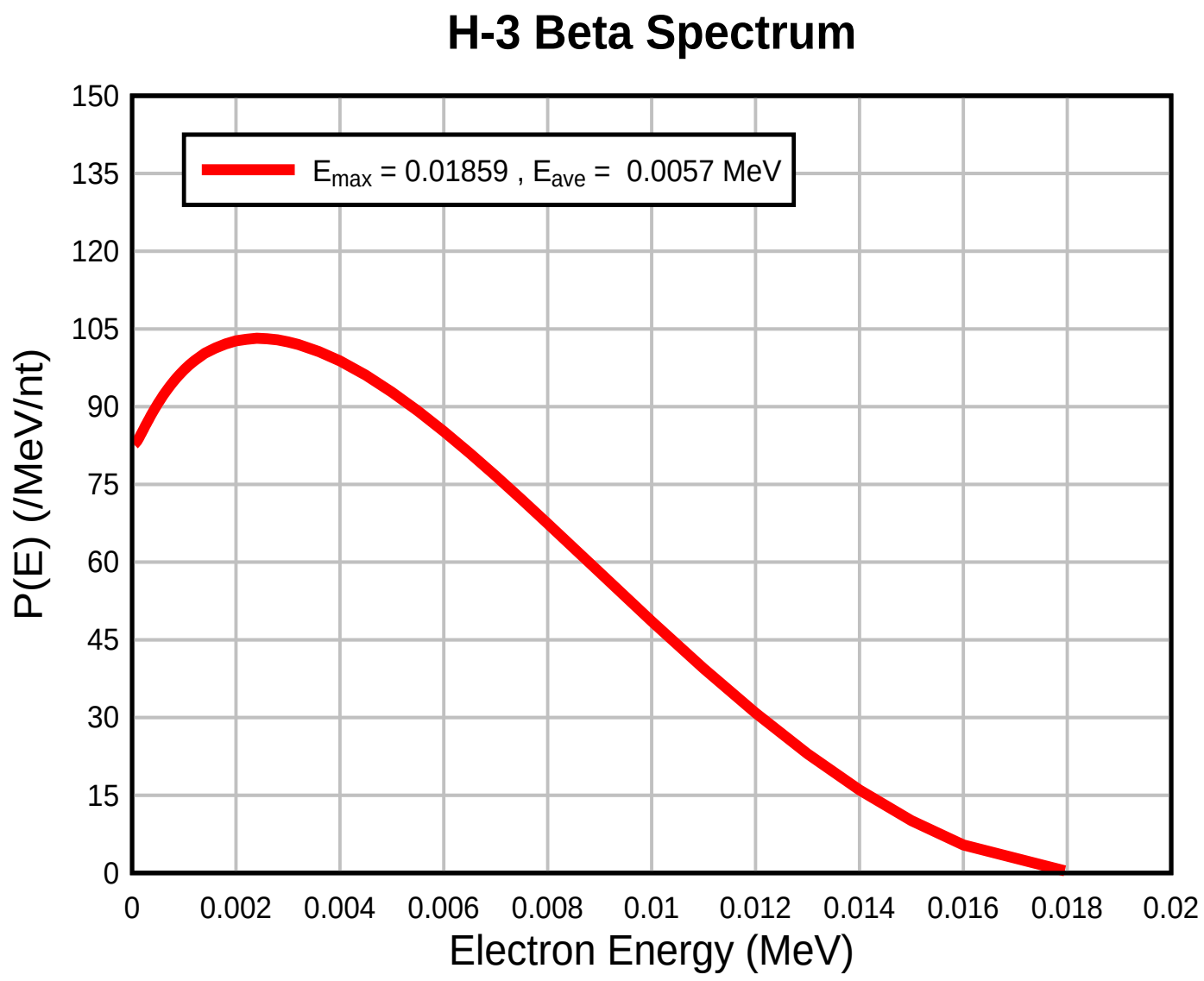

Figure 1: Spectrum, average and maximum energy of beta $^{3}$H particles

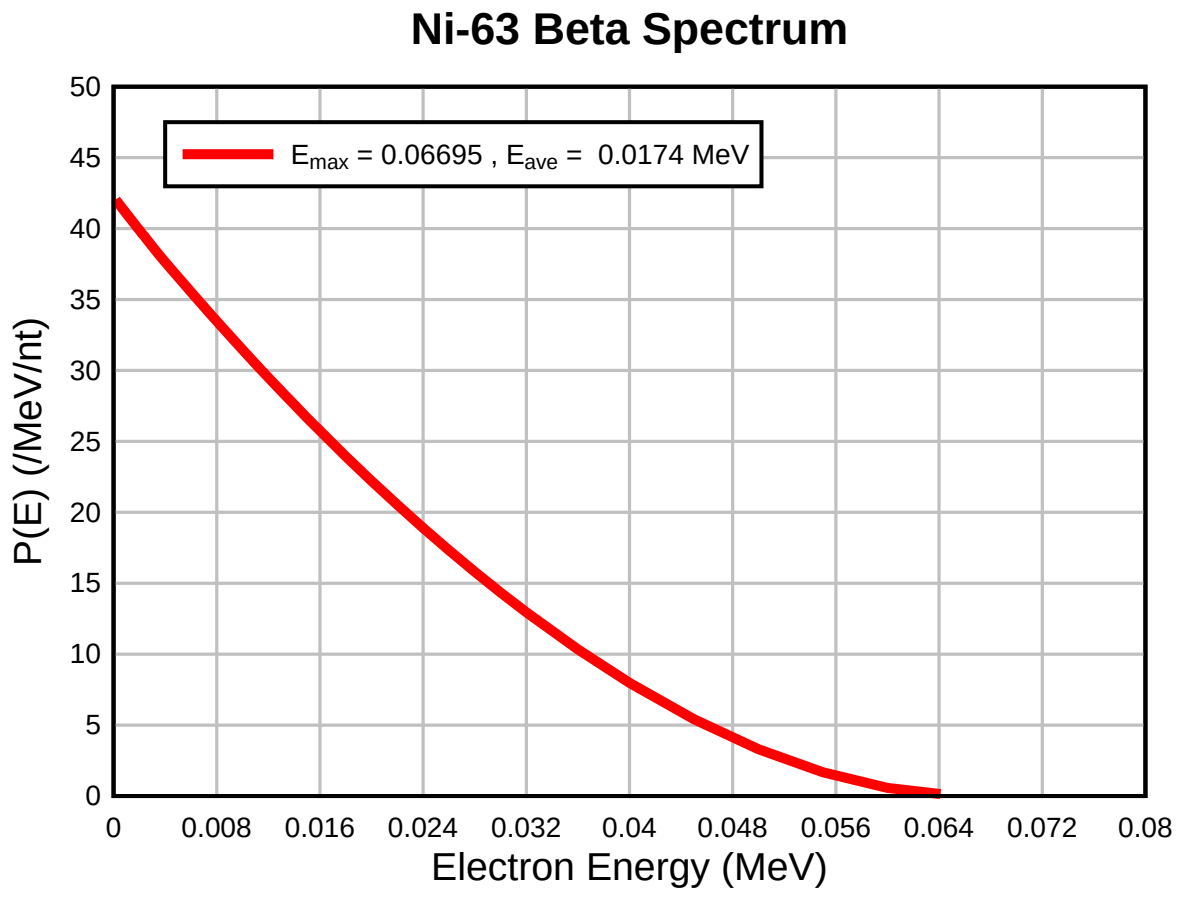

Figure 2: Spectrum, average and maximum energy of $^{63}$Ni beta particles

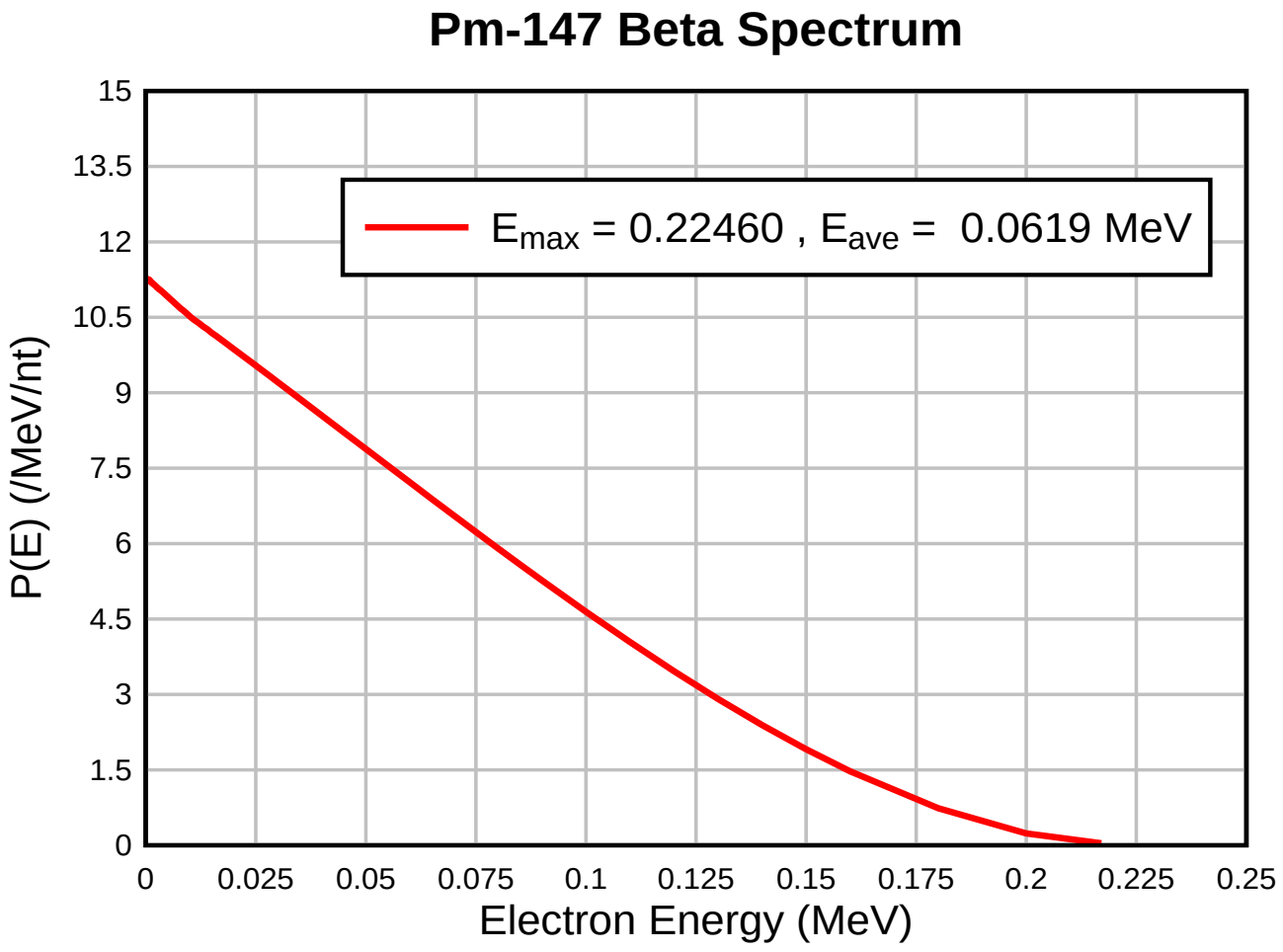

Figure 3: Spectrum, average, and maximum beta energy of $^{147}$ Pm

## 2-2- coupling efficiency ($\eta_{Couple}$)

The efficiency shows the number of beta particles entering the semiconductor and the probability of collecting the electron-hole pair is called Coupling efficiency ($\eta_{Couple}$) and is defined according to equation (11).

$$\eta_{Couple} = (1-\eta_{BSE})C_{\beta} \quad (11)$$

Where, $C_{\beta}$ is the probability of electron-hole collection and the backscattering coefficient of beta particles from the semiconductor surface. The $\eta_{BSE}$ coefficient is considered as an evaluation criterion in choosing a semiconductor as a betavoltaic cell converter. The backscattering coefficient of beta particles from the semiconductor is defined as a fraction of the electrons reflected from the surface of the semiconductor to the total number of beta particles incident on the semiconductor. The reflection yield or backscattering coefficient ($\eta_{BSE}$) is defined as equation (12).

$$\eta_{BSE} = \frac{N_{BSE}}{N_B} \quad (12)$$

Where, $N_{BSE}$ is the number of particles reflected from the surface and $N_B$ is the total number of particles reaching the semiconductor[11]. Various analytical relations have been expressed to calculate the backscattering coefficient, one of the best of them is the experimental formula of Staub in the form of equation (13)

$$\eta_0(Z,E) = \beta\left[1 - exp(-6.6\times10^{-3}\beta^{-\frac{5}{2}}Z)\right] \\ ,\beta = 0.40 + 0.065 ln(E) \tag{13}$$

where Z is the atomic number of the target element $Z \geq 4$ and E is the kinetic energy of the electron in terms of electron volts in $0.5 \leq E \leq 100keV$ the acceptable energy range[12]. To use equation (13) in compound semiconductors, the element's atomic number close to its effective atomic number is used. The effective atomic number can be calculated by equation (14).

$$Z_{eff} = \frac{\sum_{i=1}^{L}\left(\frac{w_i}{A_i}\right)Z_i^2}{\sum_{i=1}^{L}\left(\frac{w_i}{A_i}\right)Z_i} \tag{14}$$

Where, L is the number of elements in the semiconductor composition, $W_i$ is the weight fraction of the i-th element, $A_i$ is the atomic weight of the i-th element, and $Z_i$ is the atomic number of the i-th element[13].

The probability of collecting electron holes produced inside the depletion region due to the interaction of beta particles with the semiconductor is considered to be 100%. In this area, the electric field separates the holes from each other at the speed of the electron. The probability of the collection of electron holes created outside the depletion region is less than one because the electron-hole pair must penetrate into the depletion region. The collection probability in this part depends on the distance of the minority carriers from the depletion region and depletion length. The probability of collection in p and n areas is obtained from equation (15)[14].

$$C_\beta = 1 - \tanh\left(\frac{d}{L_d}\right) \tag{15}$$

Where, d is the distance from the depletion region and $L_d$ the propagation length of the minority carriers in the semiconductor; which may not necessarily be the same on both sides of the depletion area. Based on equation (15), the probability of collecting electron-holes that are created outside the depletion region and their distance to the depletion region is greater than the depletion length of the minority carriers is not considered; Therefore, semiconductors with higher carrier mobility have larger minority carrier propagation lengths and will have better efficiency. This feature is determined as one of the semiconductor selection factors.

The range of beta particles should be proportional to this active length of the converter and the energy stored from them in the semiconductor in this part, especially in the depletion region, should be maximum possible. One of the most accurate relationships for calculating the range of a beta particle with energy E in matter is Kanaya and Ekayama's equation according to equation (16).

$$R(cm) = \frac{2.76\times10^{-11} AE^{\frac{5}{3}}(1+0.978\times10^{-6}E)^{\frac{5}{3}}}{\rho Z^{\frac{8}{9}}(1+1.958\times10^{-6}E)^{\frac{5}{3}}} \tag{16}$$

Where, Z is the atomic number of the target, A is the atomic weight of the target in grams,ρ mass density in $\frac{gr}{cm^3}$ , and E is the incident energy in electron volts (eV) [15].

The ionization or excitation dissipation energy rate when beta particles pass through a semiconductor material is obtained from equation (17).

$$\frac{dE}{ds} = -\frac{78500}{E}\frac{Z\rho}{A} ln\left(\frac{1.66E}{J}\right) \quad (17)$$

Where, Z is the atomic number, A is the atomic weight, and ρ is mass density, S is the distance from the path of the incoming electron, E is the electron energy, and J is the average energy to produce an electron-hole pair[16]. For more accurate calculation of the energy loss of the energy spectrum of beta particles in semiconductors, powerful Monte Carlo codes such as Geant4 or MCNP can be used. According to equation (16), the penetration of beta particle radiation in semiconductor material depends on its density. The penetration depth of radiation is shorter in materials with lower density, and therefore a greater thickness of that material is needed to absorb incident radiation.

For example, the density of semiconductor C(diamond) is approximately 1.5 times the density of Si, the thickness required under the same conditions to absorb the incident radiation will be less and abtwo-thirdsirds of the thickness of silicon.The penetration depth of beta particles and the depletion length of the minority carriers should be well matched, so that the energy profile stored per unit length of the semiconductor in its active region is the maximum possible;Therefore, the appropriateness of the penetration depth (range) of beta particles in the semiconductor and the stored energy profile in the active area of the converter can be considered as a factor in choosing the appropriate semiconductor.

When the energy of incident beta particle is improved, the proportion of radiation loss of energy will increase. loss of energy refers to energy loss when beta particle collides with the nucleus of the target substance atom under Coulomb force. The incident beta particle interacts with the nucleus can change the speed and direction of electrons with emitting electromagnetic waves. This radiation loss of energy is also called bremsstrahlung loss. Equation (18) is used to estimate the bremsstrahlung yield in semiconductors.

$$Y \cong \frac{6\times10^{-4}ZE}{1+6\times10^{-4}ZE} \quad (18)$$

Where, Y is the radiation yield, Z is the effective atomic number of the semiconductor absorber, and E is the electron energy in MeV[17]. Because beta particles have a continuous spectrum of energy up to a maximum value, to compare the radiation yield, the average energy or the maximum energy of beta particles was considered. Equation (18) shows that the radiation yield of bremsstrahlung is directly related to the increase in atomic number and energy of beta particles;Therefore, higher beta particle energy leads to more bremsstrahlung production in the semiconductor. the lower the incident electron energy and the higher the atomic number of the target material contribute to the greater the backscatter loss of beta particles on the surface of the semiconductors.

## 2-3- Semiconductor efficiency($\eta_{semi}$)

Semiconductor efficiency ($\eta_{semi}$) is defined as equation (19).

$$\eta_{semi} = \frac{qV_{oc}FF}{w} \quad (19)$$

Where, q is the electron charge, $V_{oc}$ is open circuit voltage of the betavoltaic cell, w is the average energy to produce an electron-hole pair, and FF is the filling factor. The relationship between the band gap and the output voltage is defined as equation (20)

$$qV_{oc} = \eta_{dp}E_g \quad (20)$$

According to the equations (19) and (20), it can be obtained from the equation (21) for the semiconductor efficiency.

$$\eta_{semi} = \frac{\eta_{dp} E_g FF}{w} \quad (21)$$

$\eta_{pp}$ , The efficiency of electron-hole pair production is defined by equation (22).

$$\eta_{pp} = \frac{E_g}{w} \quad (22)$$

The dependence of the bandgap on the pair production efficiency is in the form of equation (23).

$$\eta_{pp} = \frac{E_g}{2.8E_g + 0.5} \quad (23)$$

The semiconductor efficiency of equation (21) is rewritten as equation (24)[18].

$$\eta_{semi} = \eta_{dp} \eta_{pp} FF \quad (24)$$

The importance of knowing the electron-hole pair generation efficiency is that according to the equation (24) and that the values of FF and $\eta_{dp}$ are smaller than one; The maximum semiconductor efficiency is equal to $\eta_{pp}$ ; Therefore, the efficiency of a betavoltaic cell can never be exceeded $\eta_{pp}$ .

$$\eta_{semi} \le \eta_{pp} \Rightarrow \eta_{semi} \le \frac{E_g}{w} \quad (25)$$

The maximum efficiency or in other words the final limit of efficiency is independent of the power and energy of the Beta source; Therefore, in choosing the optimal semiconductor, the electron-hole pair production efficiency should be considered as one of the important variables in choosing the semiconductor. Considering the importance of electron-hole pair production efficiency, maximum semiconductor coupling efficiency, in this research, the combined efficiency for semiconductor materials is defined as equation (26); And it has been placed as an evaluation criterion in the selection of semiconductor materials.

$$\eta_{cs} = (1 - \eta_{BSE}) \eta_{pp} \quad (26)$$

According to the above content, the efficiency of the photovoltaic battery can be rewritten in the form of coefficients involved in it in the form of equation (27):

$$\begin{aligned} \eta_{totall} &= \eta_{\beta} \eta_{couble} \eta_{semi} \\ &= \eta_{\beta} C_{\beta} (1 - \eta_{BSE}) \eta_{pp} \eta_{dp} FF \\ &= \eta_{\beta} C_{\beta} \eta_{cs} \eta_{dp} FF \end{aligned} \quad (27)$$

**2-4- Effect of temperature on betavoltaic cell efficiency**

The band gap width of semiconductors is somewhat dependent on temperature, and the density of intrinsic carriers of semiconductors is also a function of temperature. In general, the increase in temperature causes a decrease in these parameters, and as a result, the efficiency of the betavoltaic battery decreases[19]; therefore, semiconductor materials with minimal changes in efficiency with increasing temperature are more suitable choices.

**2-5- Semiconductor radiation damage threshold**

According to the energy of beta particles, a type of semiconductor should be selected that can withstand the radiation power corresponding to the beta source. The radiation damage threshold of the semiconductor should be considered as a parameter in choosing the appropriate semiconductor. A semiconductor that has a higher radiation tolerance should be used and its radiation damage energy threshold should be higher than the maximum energy of the beta particle spectrum as much as possible. Especially, this matter should be considered in the active region of the semiconductor (depletion region plus the depletion length of the minority carriers). By using the Rutherford scattering theory, the energy transferred to the atom is determined by nuclear collisions. For electrons, the maximum energy transferred is

$$T_m = \frac{2m_e}{M}\left[\frac{(E + 2m_e c^2)E}{m_e c^2}\right] \quad (28)$$

Where me is the mass of an electron, M is the mass of the atom to be displaced, c is the speed of light, and E is the energy of the incoming particle. The threshold energy of the particle Eth occurs when Tm = Ed. The threshold energy is defined as the minimum energy required for displacing an atom in the crystal [15].

$$E_d = 2E_{th}\left[\frac{\left(E_{th} + 2m_e c^2\right)}{Mc^2}\right] \Rightarrow E_{th} = -m_e c^2 + \frac{1}{2}\sqrt{4(m_e c^2)^2 + 2E_d Mc^2} \quad (29)$$

The minimum energy for atomic displacement is directly related to the bond strength of the semiconductor. It is approximately twice the bond energy. The minimum energy for atomic displacement ($E_d$) is

$$E_d(eV) = 1.7926 \times \left(\frac{1}{a_0(nm)}\right)^3 \quad (30)$$

Where $a_0$ is the Lattice constant in nanometers (nm) [20].

**3- Materials and methods**

According to the theoretical foundations stated in section 2, the factors influencing the performance of a semiconductor piece for an emission-electrical converter, the main criteria for choosing a suitable semiconductor include: the backscattering coefficient of beta particles from the semiconductor, the electron-hole pair generation efficiency, electronic characteristics and characteristics, threshold Radiation damage, Bremsstrahlung radiation production yield, stopping power and penetration of beta particles in semiconductors, physical characteristics and temperature tolerance and finally accessibility and manufacturing are considered.

Based on these criteria and by comparing with silicon semiconductor, conventional semiconductors have been quantitatively evaluated. The calculations related to each evaluation criterion are performed according to the relevant relationships and the results are expressed in the form of graphs and tables. In the end, according to the comparison of all the evaluation criteria, the optimal selection of the semiconductor with the appropriate radioisotope source is done.

Considering that the semiconductors have an effective atomic number less than 14 and a band gap higher than 1.12 eV at room temperature, 10 semiconductors, diamond, 2H-SiC, 3C-SiC, 4H-SiC, AlN, MgO, B4C was selected and evaluated. The specifications of these 10 semiconductors are given in Table (2)[20] .

Table 2 Characteristics of selected semiconductors

| **Semiconductor** | **atomic number** | **density** | **Average atomic weight $A_{av}$** | **Effective atomic number $Z_{eff}$** | **band gap** | **Type of band gap** | **W** | **Electron mobility $\mu_e$ $(\frac{cm^2}{Vs})$** | **Hole mobility $\mu_p$ $(\frac{cm^2}{Vs})$** |
|---|---|---|---|---|---|---|---|---|---|
| Si | Si(Z=14) | 2.33 | 28 | 14 | 1.12 | Indirect | 3.66 | 1500 | 480 |
| $\beta - B$ | B (Z=5) | 2.35 | 10.8 | 5 | 1.5 | Indirect | 4.7 | 300-10 | 2 |
| Diamond | C (Z=6) | 3.52 | 12 | 6 | 5.48 | Indirect | 13.2 | 2000 | 1600 |
| 2H-SiC | Si (Z=14)<br>C (Z=6) | 3.219 | 40.1 | 11.6 | 3.23 | Indirect | 9.544 | 900 | 150 |
| 3C-SiC | Si (Z=14)<br>C (Z=6) | 3.215 | 40.1 | 11.6 | 2.93 | Indirect | 8.704 | 800 | 40 |
| 4H-SiC | Si (Z=14)<br>C (Z=6) | 3.290 | 40.1 | 11.6 | 3.26 | Indirect | 7.28 | 1000 | 120 |
| 6H-SiC | Si (Z=14)<br>C (Z=6) | 3.215 | 40.1 | 11.6 | 3.05 | Indirect | 6.9 | 400 | 100 |
| $c - BN$ | B (Z=5)<br>N (Z=7) | 3.45 | 24.8 | 6.17 | 6.4 | Indirect | 17.6 | 500 | 500> |
| AlN | Al (Z=5)<br>N (Z=7) | 3.26 | 41 | 10.9 | 6.19 | direct | 15.3 | 300 | 14 |
| MgO | Mg (Z=12)<br>O (Z=8) | 3.58 | 40.3 | 10.4 | 7.8 | direct | 22.34 | 2 | 1 |
| $B_4C$ | B (Z=5)<br>C (Z=6) | 2.52 | 55.3 | 5.55 | 2.09 | direct | 6.352 | 1 | 2 |

## 4- Results and discussion

### 4-1- Electronic characteristics of semiconductors

In this part, electronic characteristics of semiconductors are investigated as an evaluation index.

### 4-1-1- Electron and hole mobility

As can be seen from Table (2), indirect bandgap semiconductors will have a larger propagation length than direct bandgap semiconductors. Due to the low electron and hole mobility of semiconductors, MgO, B4C, and AlN cannot be used in semiconductors as betavoltaic battery converters because due to very low electron and hole mobility, they will not have the ability to produce a suitable current, so they will not be evaluated in other factors. took Among the semiconductors 2H-SiC, 3C-SiC, 4H-SiC, 6H-SiC, and 4H-SiC have a higher energy gap and better electrical characteristics when used in betavoltaic batteries and will be studied. According to electron and hole mobility, respectively, diamond, Si, 4H-SiC, and C-BN have higher electron and hole mobility. Among them, only diamond has the ability to excite electron holes more than silicon.

### 4-1-2- Production saturation current in semiconductor

The minimum saturation current at a temperature of 300 K has been calculated using equation (6) for selected semiconductor materials according to table (3).

Table 3. Calculation of minimum saturation current in selected semiconductors

| Semiconductor | **Bandgap type** | **Bandgap (eV)** | Minimum saturation current density $\left(\frac{A}{cm^2}\right)$ |
|---|---|---|---|
| Si | Indirect | 1.12 | $2.29\times10^{-14}$ |
| C(diamond) | Indirect | 5.48 | $1.29\times10^{-87}$ |
| 4H-SiC | Indirect | 3.26 | $2.55\times10^{-50}$ |
| c-BN | Indirect | 6.4 | $4.49\times10^{-103}$ |

Table (3) shows that, semiconductors with a larger energy Bandgap have a lower production saturation minimum current. This causes them to have a larger open circuit voltage. The lowest value of saturation current corresponds to c-BN and C (diamond) semiconductors. This result can be justified due to the high energy gap in these semiconductors.

### 4-2- Backscattering coefficient of beta particles from semiconductor

In figure (4) the backscattering coefficient in terms of atomic number for energies of 5.68, 17.42, 61.92 keV average spectrum of beta isotopes $^{3}H$, $^{63}Ni$, $^{147}Pm$ in the form of radiation perpendicular to the semiconductor surface for semiconductors with atomic numbers less than 14 (silicon atomic number index) ) was extracted based on equation (20).

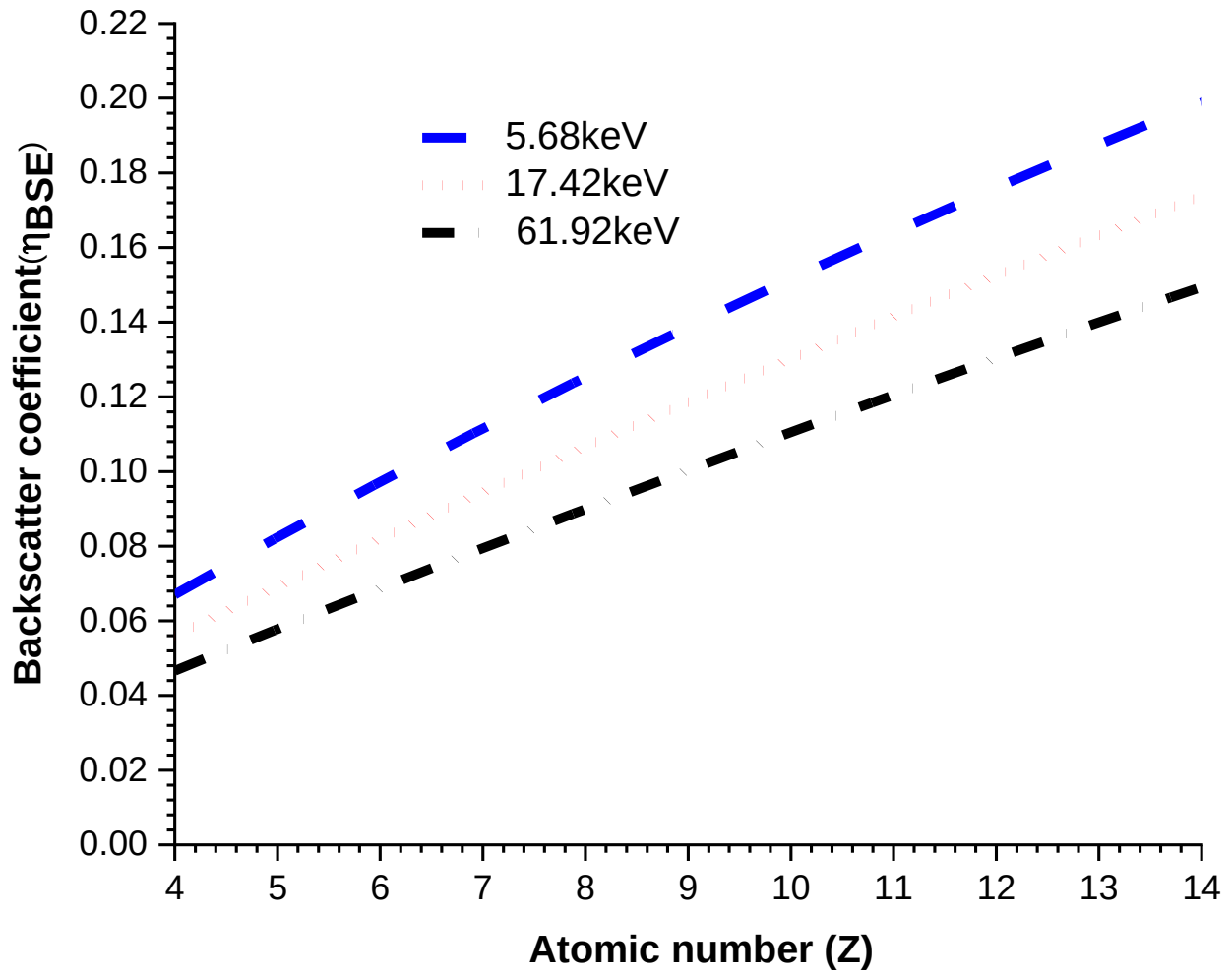

Figure 4. Backscattering coefficient for semiconductors with an effective atomic number lower than silicon for three energies: 5.68, 17.42 and 61.92 keV

figure (4) shows that under the same conditions of energy and direction, beta particles with low energy and target material with high atomic number cause the most directional loss and backscattering of particles from the material; Therefore, this point should be taken into account when choosing a semiconductor and a beta source for a betavoltaic battery. Using the Staub formula, the maximum coupling efficiency of the semiconductor to the source was calculated according to equation (11) by including the coefficient $C_\beta = 1$ based on different energies for selected semiconductor materials according to figure (6).

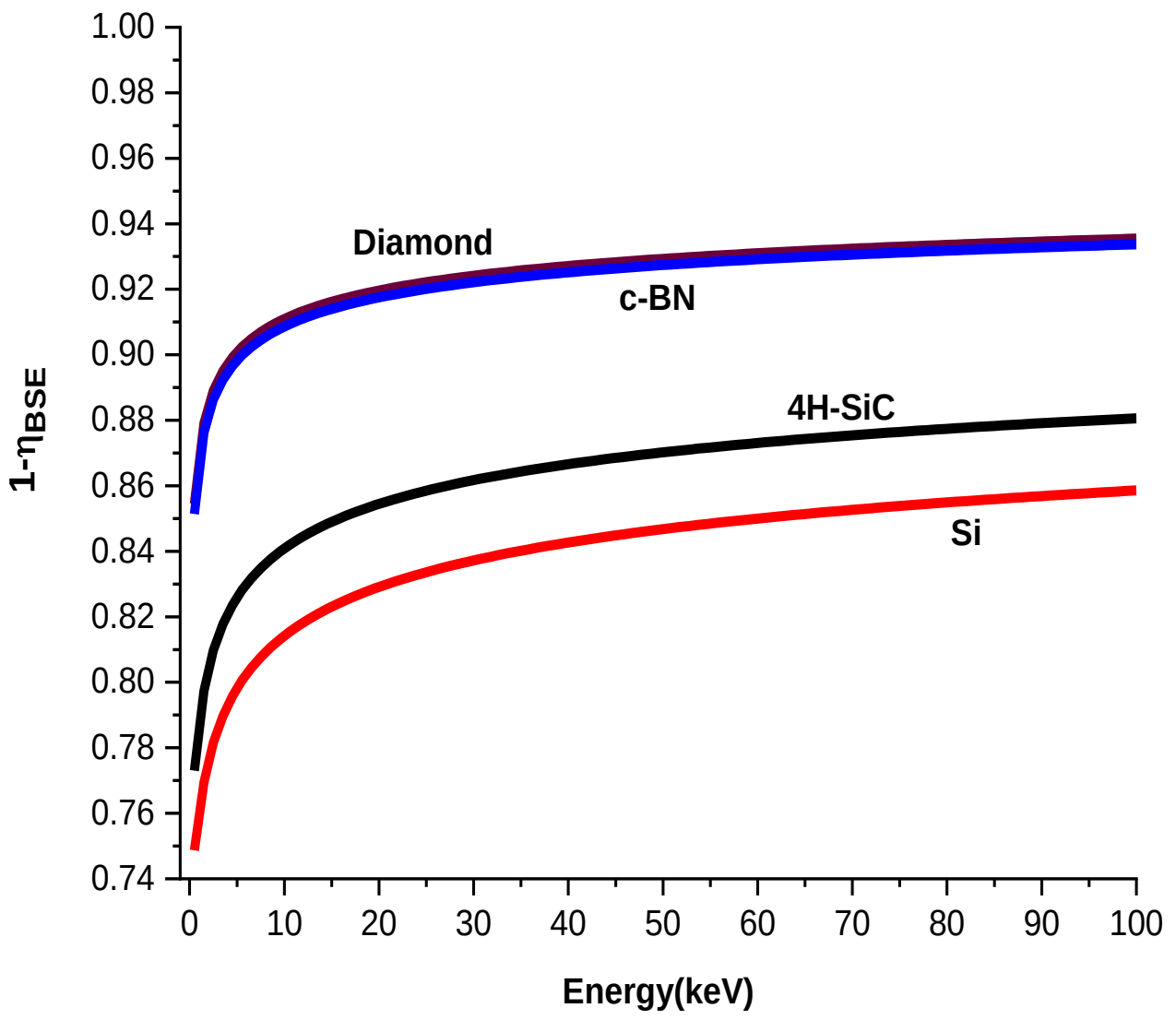

Figure 5. Calculation of the maximum coupling efficiency of the semiconductor to the beta source

According to figure (5), the maximum coupling efficiency is corresponding to Diamond, C-BN, 4H-SiC and Si respectively.

## 4-3- Paired production efficiency

In Table (4) bandgap energy, the energy required to produce electron-hole pairs and the efficiency of electron-hole pair production for selected semiconductors at 300 K temperature have been calculated. The electron-hole pair energy for semiconductors, which was not available experimentally in reference [19], have been calculated from the Klein theory equation.

Table 4. Bandgap energy, the energy required to produce an electron-hole pair and the pair production efficiency

| **Semiconductor** | **Eg(eV) At 300 Kelvin** | **W(eV)** | $\eta_{pp} = \frac{E_g}{w}$ |
|---|---|---|---|
| Si | 1.12 | 3.64 | 30.77 |
| 4H-SiC | 3.26 | 9.63 | 33.85 |
| Diamond | 5.48 | 15.84 | 34.60 |
| c-BN | 6.4 | 18.42 | 34.74 |

In figure (6), the graph of selected semiconductors is drawn according to the electron-hole pair production efficiency. According to the graph, electron-hole pair production efficiency is maximum for c-BN semiconductors.

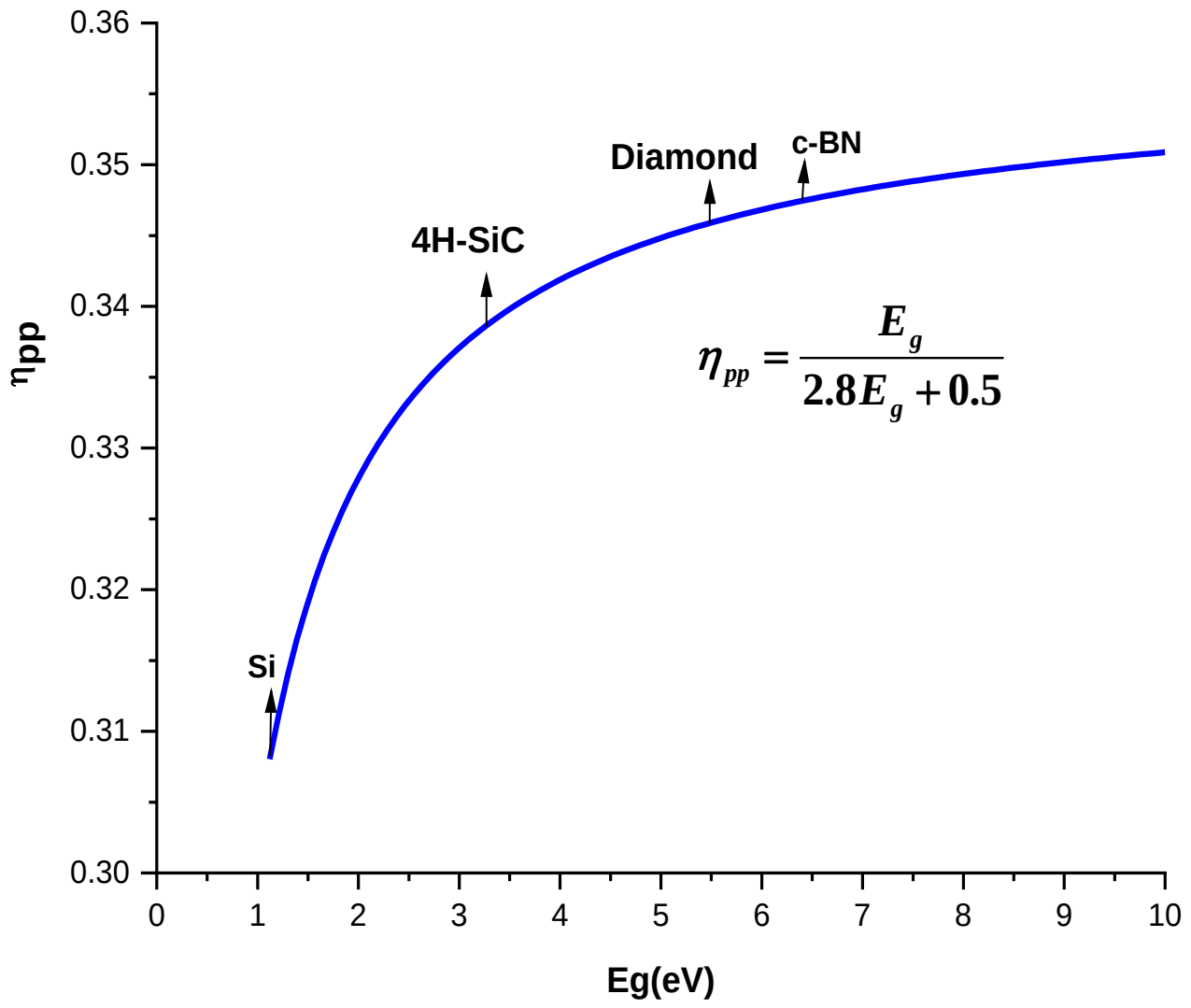

Figure 6. Pair production efficiency for selected semiconductors at temperature (300 K)

The combined efficiency ($\eta_{cs}$) for different energies for selected semiconductor materials was calculated in the form of the diagram in Figure (7).

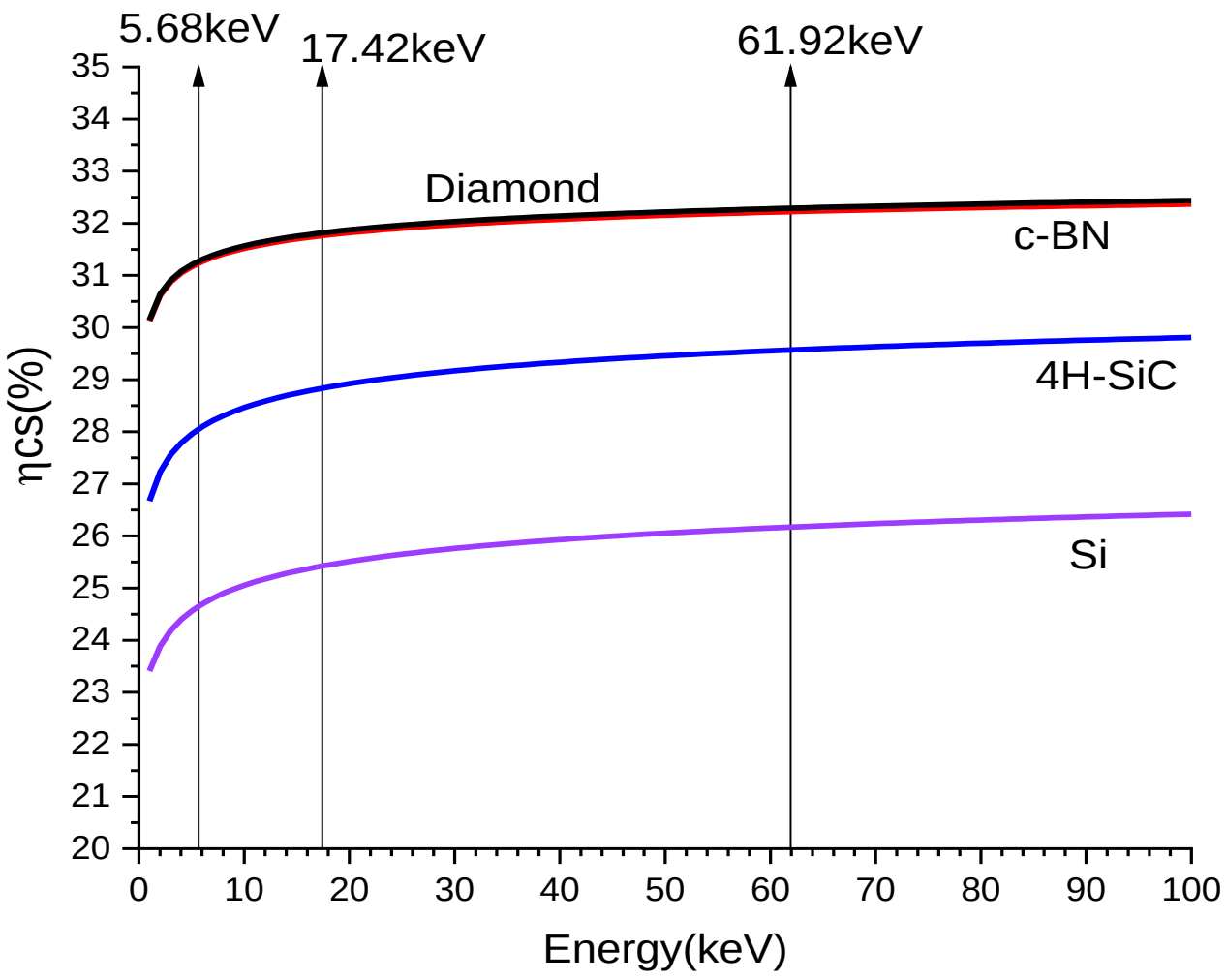

Figure 7. $\eta_{cs}$ for different energies for selected semiconductors

As can be seen, the $\eta_{cs}$ of diamond, c-BN and 4H-SiC semiconductors is greater than that of silicon. For example, for the energy of 17.42 keV, equivalent to the average energy of the $^{63}$Ni radioisotope spectrum, the efficiency of diamond and c-BN semiconductors is 1.25 times higher than the efficiency of silicon.

**4-4- Semiconductor radiation damage threshold**

According to the investigations carried out in the previous sections, three semiconductors C(diamond), 4H-SiC and c-BN are selected as suitable candidates for use as semiconductor converters in betavoltaic cells. The radiation damage threshold of selected semiconductors is according to table (5).

Table 5. Radiation damage threshold of selected semiconductors

| **Semiconductor** | **Bandgap $E_g$ (eV)** | **Displacement energy $E_d$(eV)** | **Radiation damage threshold $E_{th}$(eV)** |
|---|---|---|---|
| Si | 1.12 | 12.9 | 140 |
| C(diamond) | 3.26 | 43 | 215 |
| 4H-SiC | 5.48 | 28 | 108 |
| c-BN | 6.4 | 37.8 | 164 |

According to table (5), semiconductor C (diamond) has the highest threshold for radiation damage.

### 4-5- Range of beta particles in semiconductors

This factor is important for measuring the required semiconductor thickness and determining the dimensions of the betavoltaic cell. In figure (8), the range of beta particles with different energies in selected semiconductors is calculated using equation (16).

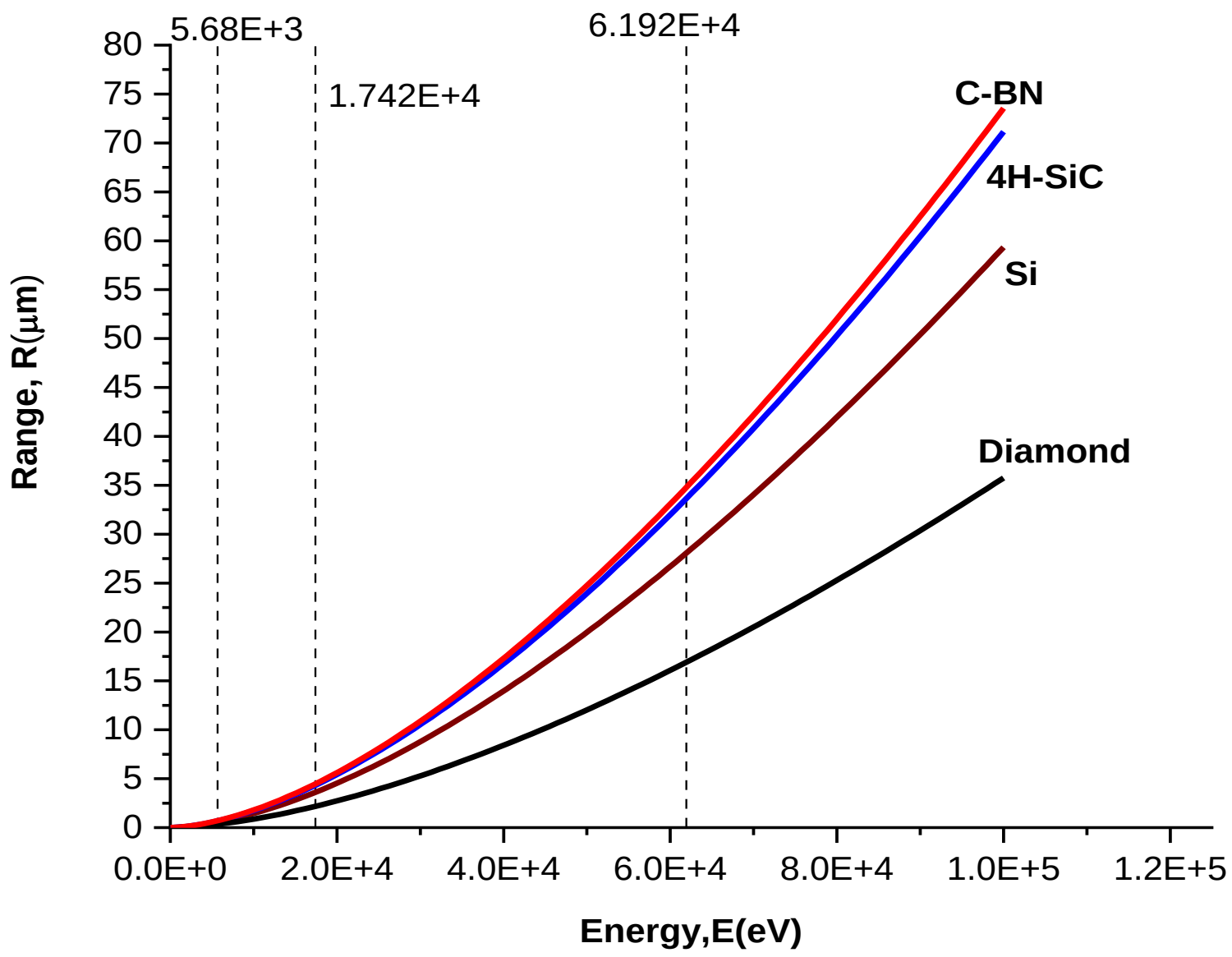

Figure 8. Beta particle range with different energies in selected semiconductors

Figure (8) shows that in order to stop beta particles Diamond, Si, 4H-SiC and C-BN, a smaller thickness of the semiconductor material is needed when used as a converter.

### 4-6- Bremsstrahlung generation yield

In figure (9), the bremsstrahlung radiation yield for selected semiconductors is calculated based on the equation (16).

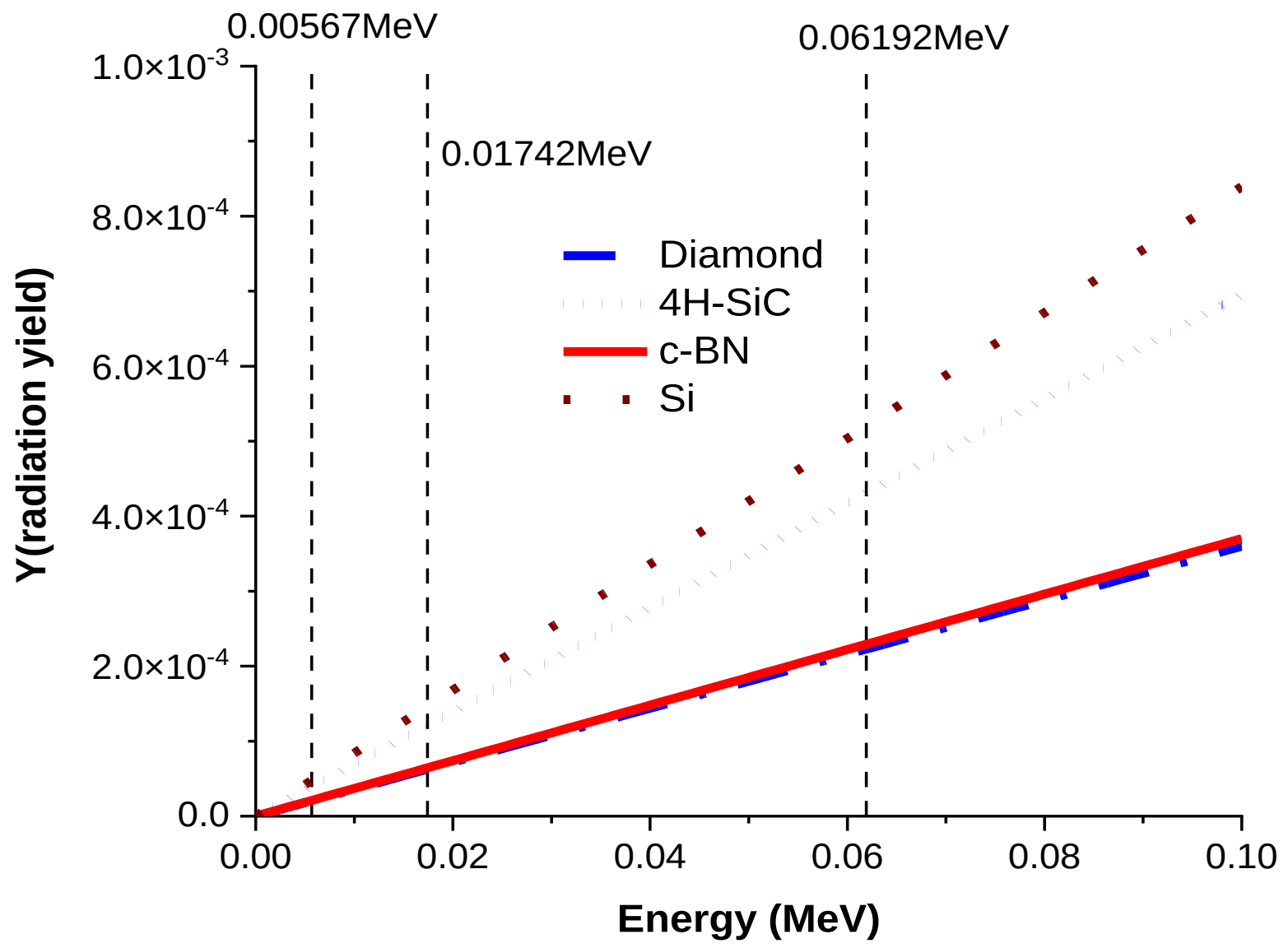

Figure 9. Bremsstrahlung radiation yield calculated for selected semiconductors

Figure (9) shows that the yield of Bremsstrahlung radiation is directly related to the increase in atomic number and energy of beta particles; Therefore, semiconductors with higher effective atomic numbers will produce more bremsstrahlung radiation. According to the above tables, C (diamond) semiconductor has the lowest Bremsstrahlung radiation yield, followed by the c-BN semiconductor, and the 4H-SiC semiconductor is ranked third.

### 4-7- Physical characteristics and temperature tolerance

The physical characteristics of the selected semiconductor materials are according to table (6).

Table 6. Physical characteristics of selected semiconductors

| Semiconductor | Lattice constant(nm) | atomic number | Density $gr/cm^3$ | melting point (K) | Hardness $kg/mm^2$ |
|---|---|---|---|---|---|
| Si | 0.543 | 14 | 2.33 | 1687 | 1150 |
| Diamond | 0.357 | 6 | 3.52 | 4100 | 7000 |
| 4H-SiC | 0.3073 | 14-6 | 3.29 | 3103 | 3980 |
| c-BN | 0.362 | 5-7 | 3.45 | 3246 | 4500 |

Table 6 shows that for selected semiconductors in terms of strength and hardness, mass density, melting temperature, and temperature tolerance. In this respect, c-BN semiconductors and 4H-SiC semiconductors are in the next ranks.

### 4-8- Accessibility, cost and construction

### 4-8-1- Silicon (Si) Semiconductor

Silicon is the first semiconductor material used as a converter. Silicon components have lower currents and are much cheaper than other semiconductor materials. The preparation process of semiconductor materials is a key relationship in the preparation of betavoltaic batteries. Monocrystalline silicon has attracted the attention of many researchers due to its low preparation cost, technological maturity, stable performance, and mass production. It is a more suitable option for the three-dimensional structure of silicon[21].

#### 4-8-2- Silicon carbide (SiC) semiconductor

The development of semiconductors with a wide energy gap has shown a good potential for use in betavoltaic batteries. However, they have a more difficult manufacturing and preparation process and have a slower speed of development and technological maturity.Silicon carbide is a wide energy gap semiconductor with high thermal stability, stable chemical properties, good electron transport performance, and radiation resistance, and is a growing technology[22].

#### 4-8-3- Diamond semiconductor (C)

The diamond semiconductor is one of the crystals with a wide band gap, which is a very hard and mechanically stable material and has a high heat capacity. Also, due to the high resistance of this material yields radiation damage, it will suffer a voltage drop later and as a result, it will have a more stable voltage.it is important to note that specialized processes, such as microwave plasma chemical vapor deposition (CVD), are required to produce diamond films. These processes involve the use of high-energy microwave radiation to create a plasma environment where carbon atoms can be deposited onto a substrate, resulting in the formation of diamond.The problem with the diamond semiconductor is that it can only do p-type doping well. Therefore, it is unsuitable for pn junction. But it is a good option in making Schottky semiconductors for use in betavoltaic batteries[23], [24].

#### 4-8-4- c-BN semiconductor

It has the capability of n and p type pollution; But it is not possible to manufacture with dimensions higher than millimeters and its technology has not reached enough maturity[25], [26]. In terms of electronic characteristics, compared to diamond semiconductor according to table (2), it is almost one quarter. Therefore, in the competition with the diamond semiconductor, it is not preferred. According to the review and the results of calculations and quantitative evaluation, the prioritization of selected options was done according to different evaluation indicators and is given in table (7).

## 5- Conclusion

In this research, based on increasing the maximum efficiency of the betavoltaic battery and the possibility of using it with $^{3}H$, $^{63}Ni$, and $^{147}Pm$ beta sources, the optimal semiconductor selection criteria are determined. and shown in table(7).

Table 7. Prioritizing options according to evaluation indicators

| index evaluation | Precedence (1) | Precedence (2) | Precedence (3) | Precedence (4) |
|---|---|---|---|---|
| Electronic characteristics of semiconductors (electron and hole movement) | diamond | Si | 4H-SiC | C-BN |
| Electronic characteristics of semiconductors (saturation current) production | c-BN | diamond | 4H-SiC | Si |
| Backscattering coefficient of beta particles from semiconductor from lowest to highest) | diamond | c-BN | 4H-SiC | Si |
| Electron-hole pair efficiency | C-BN | Diamond | 4H-SiC | Si |
| Semiconductor radiation damage threshold | Diamond | C-BN | Si | 4H-SiC |
| Semiconductor-to-source coupling efficiency | Diamond | C-BN | 4H-SiC | Si |
| Range of beta particles in semiconductor (required thickness from lowest to highest) | Diamond | Si | 4H-SiC | C-BN |
| Bremsstrahlung generation benefit from lowest to highest) | Diamond | C-BN | 4H-SiC | Si |
| Physical characteristics and temperature tolerance | Diamond | C-BN | 4H-SiC | Si |
| Availability and availability | Si | 4H-SiC | Diamond | C-BN |

These evaluation criteria include the backscattering coefficient of beta particles from the semiconductor, electron-hole pair production efficiency, electronic specifications and characteristics, radiation damage threshold, Bremsstrahlung radiation production yield, stopping power and penetration of beta particles in the semiconductor, physical characteristics and temperature tolerance, accessibility and manufacturing. , was extracted. Based on these criteria and by comparing with silicon semiconductors semiconductors are evaluated. According to the survey results and evaluation indices, Diamond, c-BN, and 4H-SiC semiconductors are selected in terms of efficiency. According to this paper for betavoltaic battery in terms of two-dimensional structures of the betavoltaic battery for Schottky bond type diamond with $^{147}$pm radioisotope, also for the 4H-SiC semiconductor with $^{63}$Ni or $^{3}$H radioisotopes and for the three-dimensional structures of betavoltaic batteries combining Si with $^{147}$Pm radioisotopes or $^{63}$Ni is recommended.

## 5- Reference